\documentclass{revtex4}
\usepackage{indentfirst}
\usepackage{graphicx}
\usepackage{mathrsfs}
\usepackage{slashed}
\usepackage{epsfig}
\begin{document}
\title{Mass effect in polarization investigation at BEPC/BES and the B-factory}
\author{JIA Hanbing}
\author{MA Bo-Qiang}
\address{School of Physics and State Key Laboratory of Nuclear Physics and Technology,\\
Peking University, Beijing 100871, China}

\begin{abstract}
We consider the annihilation process of an electron-positron pair
into a pair of heavier fermions when the initial
electron and position beams are polarized. By calculating the
polarization of the final-state particles, we discuss in detail the
effect due to the produced particle masses in the $\tau$-charm
energy region at BEPC/BES, and also compare the effect with that at
the B-factory. Such a study is useful for the design of possible
polarization investigation at the BEPC/BES facility and the
B-factory.


\end{abstract}
\maketitle

\section{Introduction}

Quantum Chromodynamics (QCD) has been considered the fundamental
theory of strong interactions, and the theoretical predictions so
far are all compatible with the phenomenological observations.
However, in many cases the accuracy of theoretical calculations and
experimental measurements is still rather low for the purpose to
seek for new physics beyond the Standard Model (SM).

For testing the reliability of the SM and searching for new physics,
it is still desirable to compare more precise predictions derived
from the theory with experimental measurements of higher accuracy.
In addition to high energy experimental projects with high
precision, lower energy facilities with high luminosity can also
address such an issue, and the upgraded BEPCII/BESIII facility is
important for high precision studies.

The BEPCII facility is a double-ring $e^+e^-$ collider running
in the $\tau$-charm energy region ($E_{\mathrm{cm}} = 2.0\sim4.6$~GeV),
and it can reach a design luminosity of
$1\times10^{33}$~cm$^{-2}$s$^{-1}$ at a center-of-mass energy of
$3.78$~GeV. With this luminosity, the BESIII detector can collect,
for example, 10 billion $J/\psi$ events in one year of
running~\cite{yellowbookofbes3,Jin:2006hd,Shen:2008zzg,Harris:2008tx}.
The comprehensive studies of $e^+e^-$ annihilation in the $\tau$-charm
threshold region can provide us with novel and unique chance to study hadronization and nonperturbative dynamics. This is very valuable to
the investigations of, for example, the structure of hadrons and the
spectrum of hadronic states. The upgraded BEPCII/BESIII can play an
important role in not only testing the SM, but also in searching for
new physics beyond the SM~\cite{LiHaibo}. There is also an optional
consideration to include the polarization of at least the incident
electrons for the future upgrade of the BEPCII/BESIII
facility~\cite{huangtao}, it is thus necessary to consider some more
detailed features of the polarized $e^+e^-$ processes.

According to QCD, the process $e^+e^-\rightarrow q\bar{q}$ (a
quark-antiquark pair) is the simplest $e^+e^-$ process that ends in
hadrons. This process is extraordinarily useful in determining the
properties of elementary particles and investigating the hadronic
structure. In this paper we compute the polarized cross section for
$e^+e^-\rightarrow \mu^+\mu^- $, to the lowest order. It is the simplest
of all QED processes, but also fundamental to the understanding of
all reactions in the $e^+e^-$ collision. It is easy to extend the
results for muon production to production of other leptons and
quarks. Thus our study is useful for further study on possible
polarization investigations at $e^+e^-$ colliders.

\section{Polarized cross section for muon-antimuon pair production}

Using the Feynman rules, we can write down the amplitude for the
polarized $e^+e^-\rightarrow \mu^+\mu^- $ process:
\begin{eqnarray*}
\bar{v}^{s'}(p')(-ie\gamma^\mu)u^s(p)(\frac{-ig_{\mu\nu}}{q^2})\bar{u}^r(k)(-ie\gamma^\nu)v^{r'}(k'),
\end{eqnarray*}
where the superscripts $s$, $s'$, $r$, and $r'$ denote the polarized states of the incoming electron, position and the outgoing muon and anti-muon pair.

The squared amplitude for this process is
\begin{eqnarray*}
|M(e^-_s(p)e^+_{s'}(p')\rightarrow \mu^-_r(k)\mu^+_{r'}(k'))|^2=\frac{e^4}{q^4}(\bar{v}^{s'}(p')\gamma^\mu u^s(p)\bar{u}^s(p)\gamma_\nu v^{s'}(p')
)(\bar{u}^r(k)\gamma_\mu v^{r'}(k')\bar{v}^{r'}(k')\gamma_\nu u^{r}(k)).
\end{eqnarray*}

We compute the polarized $e^+e^-\rightarrow \mu^+\mu^-$ cross section by using the trace technology with the addition of helicity projection operators to project out the desired left- or right-handed spinors. For a massless particle,
chirality and helicity are equivalent and Lorentz-invarient. Thus $(1+\gamma_5)/2$ becomes a right-(left-)handed
helicity projection operator for a massless particle (antiparticle), and $(1-\gamma_5)/2$ becomes a left-(right-)handed helicity projection operator for a massless particle (antiparticle). $\gamma_5$ anticommutes with $\gamma_\mu$, and thus in the electron-positron annihilation the helicities of $e^-$ and $e^+$ must be opposite to each other in order to annihilate
when the electron mass is ignored. On the other hand, for a particle with a nonvanishing mass, chirality is not well defined, but states of such a particle can still be identified by their helicities in a specific reference system. In our calculation, we take the masses of both the incident electrons and the final-state particles into account.

Let us first calculate the muon half of the squared amplitude
\begin{eqnarray*}
\bar{u}^r(k)\gamma_\mu v^{r'}(k')\bar{v}^{r'}(k')\gamma_\nu u^{r}(k).
\end{eqnarray*}

The polarization projection operator for a massive particle should be
\begin{eqnarray*}
\hat{P}(\pm)=\frac{1\pm\gamma_5\slashed{n}}{2},
\end{eqnarray*}
where $n^\mu$ is a normalized spacelike vector, $n_\mu n^\mu=-1$, and is orthogonal to the particle momentum,
$n_\mu k^\mu=0$. $\hat{P}(+)$ and $\hat{P}(-)$ project out the states with polarization in direction \textbf{n} and -\textbf{n}, respectively.

We consider the helicities of the final-state particles, so we choose
\begin{eqnarray*}
n&=&(\frac{|\textbf{k}|}{m},\frac{E}{m}\frac{\textbf{k}}{|\textbf{k}|}),\\
n'&=&(\frac{|\textbf{k}'|}{m},\frac{E}{m}\frac{\textbf{k}'}{|\textbf{k}'|}).
\end{eqnarray*}
Then $\hat{P}(h)=\frac{1+h\gamma_5\slashed{n}}{2}$ projects out the $h$-helicity ($h=\pm1$) component from the final-state muon, and similarly, $\hat{P}'(h')=\frac{1+h'\gamma_5\slashed{n'}}{2}$ projects out the $h'$-helicity ($h'=\pm1$) component from the final-state anti-muon.

The final-state muon half of $|M|^2$, for an $h$-helicity muon and an $h'$-helicity anti-muon, is then
\begin{eqnarray*}
&&\bar{u}^r(k)\gamma_\mu v^{r'}(k')\bar{v}^{r'}(k')\gamma_\nu u^{r}(k)\\& =&\sum_{\mathrm{spins}}\bar{u}(k)\gamma_\mu \hat{P}'(h') v(k')\bar{v}(k')\gamma_\nu \hat{P}(h)u(k)\\& =&\mathrm{tr}[(\slashed{k}+m)\gamma_\mu\frac{1+h'\gamma_5\slashed{n'}}{2}(\slashed{k'}-m)\gamma_\nu\frac{1+h\gamma_5\slashed{n}}{2}].
\end{eqnarray*}

Similarly, the electron half of $|M|^2$ for a $\lambda$-helicity electron and a $\lambda'$-helicity positron is then
\begin{eqnarray*}
&&\bar{v}^{s'}(p')\gamma^\mu u^s(p)\bar{u}^s(p')\gamma^\nu v^{s'}(p')\\&=&\sum_{\mathrm{spins}}\bar{v}(p')\gamma^\mu \frac{1+\lambda\gamma_5\slashed{s}}{2}u(p)\bar{u}(p')\gamma^\nu \frac{1+\lambda'\gamma_5\slashed{s'}}{2}v(p')\\&=&\mathrm{tr}[(\slashed{p'}-m_e)\gamma^\mu \frac{1+\lambda\gamma_5\slashed{s}}{2}( \slashed{p}+m_e)\gamma^\nu\frac{1+\lambda'\gamma_5\slashed{s'}}{2}],
\end{eqnarray*}
where
\begin{eqnarray*}
s=(\frac{|\textbf{p}|}{m_e},\frac{E}{m_e}\frac{\textbf{p}}{|\textbf{p}|}),\\ s'=(\frac{|\textbf{p}'|}{m_e},\frac{E}{m_e}\frac{\textbf{p}'}{|\textbf{p}'|}).
\end{eqnarray*}

Calculating the trace explicitly and dotting the electron half into the final-state muon half, we find that the squared
matrix element for polarized process $e^+e^-\rightarrow \mu^+\mu^-$ in the center-of-mass frame is
\begin{eqnarray*}
&&|M(e^-_\lambda e^+_{\lambda'}\rightarrow \mu^-_h\mu^+_{h'})|^2\\&=&\frac{e^4}{4}[(1-hh'-\lambda\lambda'+hh'\lambda\lambda')(1+\cos^2\theta)+2(h\lambda+h'\lambda'-h\lambda'-h'\lambda)\cos\theta+\frac{m^2}{E^2}(1+hh'-\lambda\lambda'-hh'\lambda\lambda')(1-\cos^2\theta)\\&&+\frac{m_e^2}{E^2}(1-hh'+\lambda\lambda'-hh'\lambda\lambda')(1-\cos^2\theta)
+\frac{m^2m_e^2}{E^4}(1+hh'+\lambda\lambda'+hh'\lambda\lambda')\cos^2\theta],
\end{eqnarray*}
where $\theta$ is the angle between the directions of the final-state muon and the incident electron.

The formula of differential cross section in the center-of-mass frame is
\begin{eqnarray*}
\frac{d\sigma}{d\Omega}=\frac{1}{2E_A2E_B|v_A-v_B|}\frac{|\textbf{p}_1|}{(2\pi)^24E_{cm}}|M(p_Ap_B\rightarrow p_1p_2)|^2.
\end{eqnarray*}
For our problem, $E_A=E_B=\frac{E_{\mathrm{cm}}}{2}=E$, $|\textbf{p}_1|=\sqrt{E^2-m^2}$, so we have
\begin{eqnarray*}
&&\frac{d\sigma}{d\Omega}(e^-_\lambda e^+_{\lambda'}\rightarrow \mu^-_h\mu^+_{h'})=\frac{1}{4E^2|v_A-v_B|}\frac{|\textbf{k}_1|}{(2\pi)^28E}|M(e^-_\lambda e^+_{\lambda'}\rightarrow \mu^-_h\mu^+_{h'})|^2\\&=&
\frac{\alpha^2}{32E^2|v_A-v_B|}\sqrt{1-\frac{m^2}{E^2}}[(1-hh'-\lambda\lambda'+hh'\lambda\lambda')(1+\cos^2\theta)+2(h\lambda+h'\lambda'-h\lambda'-h'\lambda)\cos\theta\\&&+\frac{m^2}{E^2}(1+hh'-\lambda\lambda'-hh'\lambda\lambda')(1-\cos^2\theta)+\frac{m_e^2}{E^2}(1+\lambda\lambda'-hh'-hh'\lambda\lambda')(1-\cos^2\theta)
\\&&+\frac{m^2m_e^2}{E^4}(1+hh'+\lambda\lambda'+hh'\lambda\lambda')\cos^2\theta].
\end{eqnarray*}

\section{Longitudinal polarizations of particles produced at BEPC/BES}

By definition, the longitudinal polarizations of the incident
electrons and positrons~\cite{YSTsai} are
\begin{eqnarray}
\omega_1&=&\frac{N_{1+}-N_{1-}}{N_1},\label{eq1}\\
\omega_2&=&-\frac{N_{2+}-N_{2-}}{N_2},\label{eq2}
\end{eqnarray}
where $N_1=N_{1+}+N_{1-}$ and $N_2=N_{2+}+N_{2-}$. $N_{1+}$ and $N_{1-}$ are the numbers of electrons with
positive-helicity and negative-helicity, respectively. $N_{2+}$ and $N_{2-}$ are the numbers of positrons with
positive-helicity and negative-helicity, respectively.

Let $\frac{d\sigma}{d\Omega}(\lambda\lambda')$ denote the polarized differential cross section for a $\lambda$-helicity electron and a $\lambda'$-helicity positron. Thus the total number of events in the unit solid angle is proportional to
\begin{eqnarray}\label{eq3}
N_{1+}N_{2-}\frac{d\sigma}{d\Omega}(+-)+N_{1-}N_{2+}\frac{d\sigma}{d\Omega}(-+)+N_{1+}N_{2+}\frac{d\sigma}{d\Omega}(++)+N_{1-}N_{2-}\frac{d\sigma}{d\Omega}(--).
\end{eqnarray}
When both the electron and positron beams are polarized,
substituting Eq.~(\ref{eq1}) and Eq.~(\ref{eq2}) into
Eq.~(\ref{eq3}) we obtain the differential cross section as follows
\begin{eqnarray*}
\frac{1+\omega_1\omega_2}{4}[\frac{d\sigma}{d\Omega}(+-)+\frac{d\sigma}{d\Omega}(-+)]+\frac{\omega_1+\omega_2}{4}
[\frac{d\sigma}{d\Omega}(+-)-\frac{d\sigma}{d\Omega}(-+)]
\\+\frac{1-\omega_1\omega_2}{4}[\frac{d\sigma}{d\Omega}(++)+\frac{d\sigma}{d\Omega}(--)]+
\frac{\omega_1-\omega_2}{4}[\frac{d\sigma}{d\Omega}(++)-\frac{d\sigma}{d\Omega}(--)].
\end{eqnarray*}

Let $(h,h')$ denote the differential cross section with the helicities of the final-state muon and anti-muon being $h$ and $h'$, respectively. Thus the degree of longitudinal polarization of the final-state muon is
\begin{eqnarray}\label{eq4}
P=\frac{(h=1,h'=1)+(h=1,h'=-1)-(h=-1,h'=1)-(h=-1,h'=-1)}{(h=1,h'=1)+(h=1,h'=-1)+(h=-1,h'=1)+(h=-1,h'=-1)}=\frac{2(\omega_1+\omega_2)\cos\theta}{D},
\end{eqnarray}
where
\begin{eqnarray*}
D&=&(1+\omega_1\omega_2)(1+\cos^2\theta)+\frac{m^2}{E^2}(1+\omega_1\omega_2)(1-\cos^2\theta)+\frac{m_e^2}{E^2}(1-\omega_1\omega_2)(1-\cos^2\theta)+\frac{m_e^2}{E^2}\frac{m^2}{E^2}(1-\omega_1\omega_2)\cos^2\theta.
\end{eqnarray*}
Just substituting the corresponding final-state fermion mass for $m$, Eq.~(\ref{eq4}) is also valid for other $e^+e^-\rightarrow f\bar{f}$ processes,
where $f\bar{f}$ represents fermion pairs including $\tau^+\tau^-$ lepton pair and quark-antiquark pairs of which the production thresholds are below the center-of-mass energy.

When the masses of both the final-state fermion and the incident electron are ignored, Eq.~(\ref{eq4}) becomes
\begin{eqnarray}\label{eq5}
P_{m,m_e\rightarrow0}=\frac{2(\omega_1+\omega_2)\cos\theta}{(1+\omega_1\omega_2)(1+\cos^2\theta)}.
\end{eqnarray}

\begin{figure}
  \includegraphics[width=0.4\textwidth]{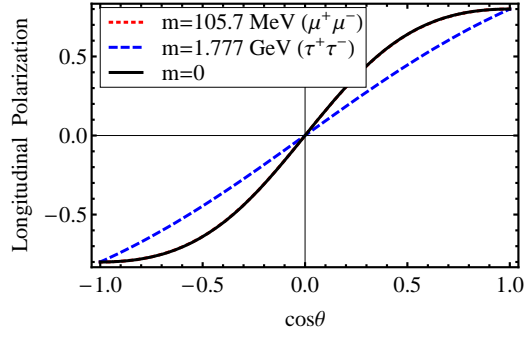}\\
  \caption{The longitudinal polarizations of the
final-state leptons produced at BEPC/BES ($E$=2.087~GeV).  In the
figure the dotted curve is in overlap with the solid curve. }\label{pic1}
\end{figure}

\begin{figure}
  \includegraphics[width=0.4\textwidth]{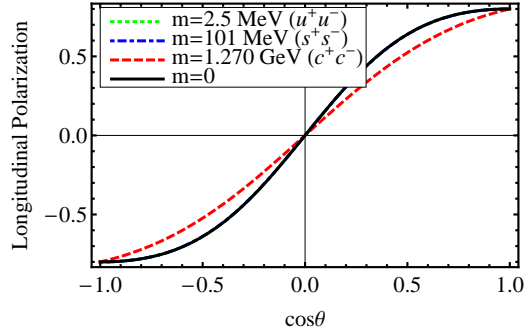}\\
  \caption{ The longitudinal polarizations of the
final-state quarks produced at BEPC/BES ($E$=2.087~GeV).
  In the figure the dotted and dot-dashed curves
  are in overlap with the solid curve.}\label{pic2}
\end{figure}

At BEPC/BES, where $E=2.087$~GeV, we consider the case that
only the electron beam is polarized while the positron beam is
unpolarized. Substituting $\omega_1=80\%$ and $\omega_2=0$ into
Eq.~(\ref{eq4}) and Eq.~(\ref{eq5}), and also substituting the
masses~\cite{Nakamura:2010zzi} of the incident electron and the
final-state fermions into Eq.~(\ref{eq4}), we can plot the curves of
final-state fermion longitudinal polarizations $P$ and
$P_{m,m_e\rightarrow0}$ versus $\cos\theta$ as shown in
FIG.~\ref{pic1} and FIG.~\ref{pic2}.

It can be seen from the figures that the longitudinal polarization curves $P$ and $P_{m,m_e\rightarrow0}$ are almost overlapped when the final-state particles are $\mu^+\mu^-$ or $u\bar{u}$. This indicates that ignoring the masses of the initial and final state particles has no significant effect on the final-state fermion longitudinal polarization in this case; when the final-state particles are $\tau^+\tau^-$ or $c\bar{c}$, the curve $P_{m,m_e\rightarrow0}$ clearly separates from the curve $P$, and this indicates that ignoring the masses of the initial and final state particles has a significant effect on the final-state fermion longitudinal polarization in this case.

When $\cos\theta=0.5$, the deviation caused by ignoring the masses of the initial and final state particles for various fermion pair productions, i. e. $\Delta P_{m,m_e\rightarrow0}=P_{m,m_e\rightarrow0}-P$, is shown in TABLE~\ref{tab1}.
\begin{table}
  \centering
  \caption{The deviation caused by ignoring the masses of the initial and final state particles for various fermion pair productions ($E$=2.087~GeV, $\cos\theta=0.5$).}\label{tab1}
\begin{tabular}{|c|c|c|}
  \hline
   & $m$ (MeV) & $\Delta P_{m,m_e\rightarrow0}$ \\\hline
  $\mu^+\mu^-$ & 105.7 & $9.83509\times10^{-4}$ \\
  $\tau^+\tau^-$ & 1777& $1.94004\times10^{-1}$\\
  $u\bar{u}$ &2.5 & $5.74039\times10^{-7}$ \\
  $d\bar{d}$ & 5.0 & $2.22709\times10^{-6}$ \\
  $s\bar{s}$& 101 & $8.98112\times10^{-4}$ \\
  $c\bar{c}$  &1270 &$1.16347\times10^{-1}$ \\
  \hline
\end{tabular}
\end{table}

When taking the final-state particle masses into account while ignoring the incident electron mass, Eq.~(\ref{eq4}) becomes
\begin{equation}
P_{m_e\rightarrow0}=\frac{2(\omega_1+\omega_2)\cos\theta}{(1+\omega_1\omega_2)(1+\cos^2\theta)+\frac{m^2}{E^2}(1+\omega_1\omega_2)(1-\cos^2\theta)}.
\end{equation}
When $\cos\theta=0.5$, the deviation caused by this approximation, i. e. $\Delta P_{m_e\rightarrow0}=P_{m_e\rightarrow0}-P$, is shown in TABLE~\ref{tab2}.
\begin{table}
  \centering
  \caption{The deviation caused by ignoring the incident electron mass for various fermion pair productions ($E$=2.087~GeV, $\cos\theta=0.5$).}\label{tab2}
\begin{tabular}{|c|c|c|}
  \hline
   & $m$ (MeV) & $\Delta P_{m_e\rightarrow0}$ \\
   \hline
  $\mu^+\mu^-$ & 105.7 & $2.29701\times10^{-8}$ \\
  $\tau^+\tau^-$ & 1777& $1.38814\times10^{-8}$\\
  $u\bar{u}$ &2.5 & $2.30212\times10^{-8}$ \\
  $d\bar{d}$ & 5.0 & $2.30211\times10^{-8}$ \\
  $s\bar{s}$& 101 & $2.29746\times10^{-8}$ \\
  $c\bar{c}$  &1270 &$1.73142\times10^{-8}$ \\
  \hline
\end{tabular}
\end{table}

\section{Longitudinal polarizations of particles produced at the B-factory}

\begin{figure}
  \includegraphics[width=0.4\textwidth]{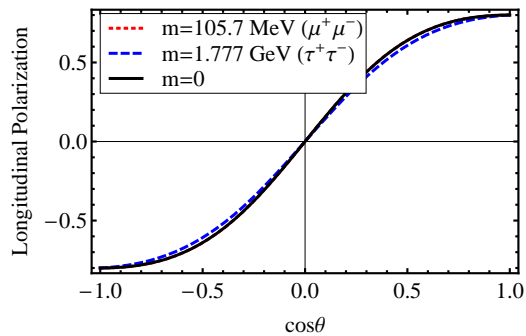}\\
  \caption{The longitudinal polarizations of the final-state leptons produced at the B-factory ($E$=6.0~GeV). In the figure the dotted curve
 is in overlap with the solid curve.}\label{pic3}
\end{figure}

\begin{figure}
  \includegraphics[width=0.4\textwidth]{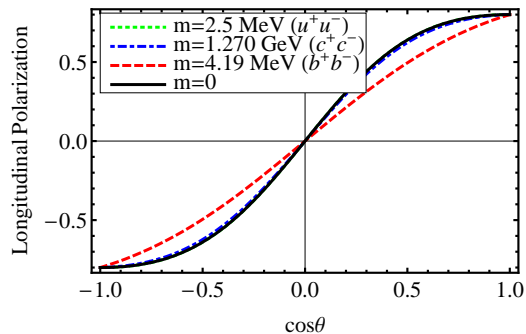}\\
  \caption{The longitudinal polarizations of the final-state quarks produced at the B-factory ($E$=6.0~GeV). In the figure the dotted and dot-dashed curves
  are in overlap with the solid curve.}\label{pic4}
\end{figure}

Then we consider the case at the B-factory. As discussed above, substituting the B-factory energy $E=6.0$~GeV for BEPC/BES energy $E=2.087$~GeV, and including the case of $b\bar{b}$ production, we can plot the curves of the final-state fermion longitudinal polarization $P$ and $P_{m,m_e\rightarrow0}$ versus $\cos\theta$ as shown in FIG.~\ref{pic3} and FIG.~\ref{pic4}.

At the B-factory, as shown in the figures, only the mass of $b$
quark has significant effect on the final-state fermion longitudinal
polarization, while the masses of $\tau$ lepton and $c$ quark are
negligible.

The deviation caused by ignoring the masses of the initial and final state
particles for various fermion pair productions at the B-factory,
i.e.,  $\Delta P_{m,m_e\rightarrow0}=P_{m,m_e\rightarrow0}-P$, is
shown in TABLE~\ref{tab3}.
\begin{table}
  \centering
  \caption{The deviation caused by ignoring the masses of the initial and final state particles for various fermion pair productions ($E$=6.0~GeV, $\cos\theta=0.5$).}\label{tab3}
\begin{tabular}{|c|c|c|}
  \hline
   & $m$ (MeV) & $\Delta P_{m,m_e\rightarrow0}$ \\\hline
  $\mu^+\mu^-$ & 105.7 & $1.19154\times10^{-4}$ \\
  $\tau^+\tau^-$ & 1777& $3.19984\times10^{-2}$\\
  $u\bar{u}$ &2.5 & $6.94519\times10^{-8}$ \\
  $d\bar{d}$ & 5.0 & $2.69452\times10^{-7}$ \\
  $s\bar{s}$& 101 & $1.08795\times10^{-4}$ \\
  $c\bar{c}$  &1270 &$1.67539\times10^{-2}$ \\
  $b\bar{b}$  &4190 &$1.44875\times10^{-1}$ \\
  \hline
\end{tabular}
\end{table}

\section{Conclusion}

In conclusion, we study the mass effect for the polarized
electron-positron annihilation processes at the BEPC/BES facility
and the B-factory. From our study, we find that the masses of the
final state fermion pairs may have some effects in the higher
precision measurements of these facilities with high luminosity.
Thus our study is useful for the design of possible polarization
investigation at these electron-positron colliders, and also for the
physical programs involving the production of heavy flavored
hadrons. The super-tau-charm factory has advantage for the study of
$\tau^+\tau^-$ and $c\bar{c}$ polarization effects compared with
that at B-factory, especially, observables for CP violation in
$\tau$ decays can be well defined with polarized electron/positron
beams at super-tau-charm factory. More studies are still needed for
making clear the influences due to the mass effect in various
processes for future super-tau-charm factory.

{\bf Acknowledge} This work is partially supported by National Natural
Science Foundation of China (Grants No.~11021092, No.~10975003,
No.~11035003, and No.~11120101004) and by the Research Fund for the
Doctoral Program of Higher Education (China).

\end{document}